\documentclass[10pt,a4paper]{article}
\usepackage{opex3}
\usepackage{graphics}
\usepackage{graphicx}
\usepackage{amssymb}
\usepackage[colorlinks=true,linkcolor=blue,citecolor=blue,urlcolor=blue]{hyperref}
\usepackage{geometry}
\usepackage{keyval}
\usepackage{trig}
\usepackage{times,helvet,courier}
\usepackage{url}
\usepackage{subfigure}
\usepackage{amsmath}
\usepackage{color}
\newcommand{\mean}[1]{\langle#1\rangle}
\begin{document}

%%%%%%%%%%%%%%%%%% title page information %%%%%%%%%%%%%%%%%%
%\linenumbers
\OEtitle{Differing self-similarity in light scattering spectra:\\ A potential tool for pre-cancer detection.}
\author{Sayantan Ghosh $^{a,1}$, Jalpa Soni $^{b,2}$, Harsh Purwar $^{b,3}$, Jaidip Jagtap $^{c,4}$, Asima Pradhan $^{c,5}$, Nirmalya Ghosh $^{b,*,6}$, Prasanta K. Panigrahi $^{b,7}$}
\OEaddress{$^a$ School of Physics, University of KwaZulu-Natal, \\Private Bag X54001, Durban 4000, South Africa. \\
$^b$ Dept. of Physical Sciences, Indian Institute of Science Education and Research Kolkata (IISER-K), P.O. BCKV Campus Main Office, Mohanpur 741 252, India.\\
$^c$ Dept. of Physics, Indian Institute of Technology Kanpur (IIT-K) 208 017, India.}
%\begin{center}\underline{\color{blue}{\footnotesize $^{1}$}\href{mailto:210556397@gmail.com}{\footnotesize 210556397@gmail.com},{\footnotesize $^2$}\href{mailto:jalpa.0787@gmail.com}{\footnotesize jalpa.0787@gmail.com},{\footnotesize $^3$}\href{mailto:harsh@iiserkol.ac.in}{\footnotesize harsh@iiserkol.ac.in},{\footnotesize $^4$}\href{mailto:jaidip@iitk.ac.in}{\footnotesize jaidip@iitk.ac.in},}\\ {\footnotesize $^5$}\href{mailto:asima@iitk.ac.in}{\footnotesize asima@iitk.ac.in},{\footnotesize $^6$}\href{mailto:nghosh@iiserkol.ac.in}{\footnotesize nghosh@iiserkol.ac.in},{\footnotesize $^7$}\href{mailto:pprasanta@iiserkol.ac.in}{\footnotesize pprasanta@iiserkol.ac.in}\end{center}
\OEemail{$^1$210556397@ukzn.ac.za,$^2$jalpa.0787@gmail.com,$^3$harshpurwar@hotmail.com,}\OEemail{$^4$jaidip@iitk.ac.in,$^5$asima@iitk.ac.in,$^6$nghosh@iiserkol.ac.in,$^7$pprasanta@iiserkol.ac.in}


\begin{abstract}
The fluctuations in the elastic light scattering spectra of normal and dysplastic human cervical tissues analyzed through wavelet transform based techniques reveal clear signatures of self-similar behavior in the spectral fluctuations. Significant differences in the power law behavior ascertained through the scaling exponent was observed in these tissues. The strong dependence of the elastic light scattering on the size distribution of the scatterers manifests in the angular variation of the scaling exponent. Interestingly, the spectral fluctuations in both these tissues showed multi-fractality (non-stationarity in fluctuations), the degree of multi-fractality being marginally higher in the case of dysplastic tissues. These findings using the multi-resolution analysis capability of the discrete wavelet transform can contribute to the recent surge in the exploration for non-invasive optical tools for pre-cancer detection.
\end{abstract}

 % REPLACE WITH CORRECT OCIS CODES FOR YOUR ARTICLE

%%%%%%%%%%%%%%%%%%%%%%% References %%%%%%%%%%%%%%%%%%%%%%%%%

%%%%%%%%%%%%%%%%%%%%%%%%%%  body  %%%%%%%%%%%%%%%%%%%%%%%%%%
\section{Introduction}\label{sec:Intro}
The use of optical techniques for the study of biomedical systems is a rapidly developing field that has seen a dramatic expansion in the recent years, partly due to tremendous progress in the field of lasers, fiber optics and associated technologies. Both medicine and biotechnology require appropriate instrumentation to analyze and monitor biological systems for deviations from normality. Optical methods, due to their non-invasive nature, are providing novel approaches for medical imaging, diagnosis and therapy. Considerable efforts have been made in the recent past to use optical techniques such as fluorescence spectroscopy \cite{ramanujam2000,kortum1996,ghosh2005,ghosh2002}, Raman spectroscopy \cite{haka2005} and elastic scattering spectroscopy \cite{boustany2010} for quantitative and early diagnosis of various diseases. Several optical imaging techniques like coherence gated imaging, polarization gated imaging and diffuse optical tomography are also being actively pursued for obtaining high resolution (micron scale) images of biological objects and their underlying structure \cite{fujimoto2003,schmitt1999,hebden1997,ghosh2011,jacques2002}. 
\par
For optical diagnosis, elastic and inelastic light scattering spectra from tissues are exploited. The in-elastically scattered light (via processes like fluorescence and Raman) contain useful biochemical information about the sample that can be employed for probing subtle biochemical changes as signatures of disease progression. On the other hand, elastically scattered light from biological tissues contain rich morphological and functional information of potential biomedical importance \cite{boustany2010,choi2007,gurjar2001,kalashnikov2009,choi2008,graf2005,wax2003,perelman1998,ghosh2010,tuchin2006,ghosh2006,ghosh2001,kim2006,drezek2003,yu2008}. Both the angular and wavelength dependence of the elastically scattered light from tissue can be analyzed to extract and quantify subtle morphological changes taking place during progression of a disease \cite{kalashnikov2009,choi2008,graf2005,wax2003,perelman1998,ghosh2006,ghosh2001}, and thus may be explored as a sensitive tool for early diagnosis. This would however involve appropriate modeling of light scattering in complex random media like tissues, and the development of suitable approaches to extract/interpret the morphological information contained in the elastic light scattering signal.
\par
The spatial fluctuation of the refractive index in biological tissues arising from scatterers ranging in sizes from a few nanometers to a few micrometers give rise to elastic scattering \cite{boustany2010,tuchin2006}. The lack of sufficient knowledge about the complex dielectric fluctuations in the tissues pose a formidable problem in the exact modeling of light scattering. Nevertheless, several efforts have been made in the recent years using electromagnetic (EM) theory based approaches like Mie theory and Born approximation to model and understand the scattering process from biological tissues \cite{capoglu2009,hunter2006,xu2005,sheppard2007,wu2007}. It has also been shown that the refractive index fluctuations in biological tissues are fractal in nature  which can be used to understand the structural changes in tissues induced by diseases \cite{graf2005,hunter2006,gao2010,wax2003_1,schmitt1996}.
\par
Since the tissue morphology dependent refractive index fluctuations are recorded in the elastic scattering spectra \cite{perelman2006}, analysis of elastically scattered spectral fluctuations using sophisticated fluctuation analysis tools might facilitate extraction and quantification of subtle morphological changes associated with early stages of cancer. The scaling behavior which is generally assumed to be global (mono-fractal), has been shown to manifest in the local fluctuations in various physical processes \cite{hurst1951,mandelbrot1982} and has been characterized using Multi-Fractal De-trended Fluctuation Analysis(for example see \cite{kantelhardt2002}). Wavelet Based Multi-Fractal Detrended Fluctuation Analysis (WB-MFDFA) is one other state-of-the-art technique that can be used for extracting and quantifying the self similarity at varying length scales associated with the structural changes associated with cancer progression due to the inherent use of fractal like transformation kernels.
%Recent works \cite{perelman2006} suggest that tissue morphology dependent refractive index fluctuations are represented in the elastic light scattering spectra through their angular and spectral dependence. The structural changes though subtle in the early stages of cancer (like dysplasia), through the use of sophisticated fluctuation analysis techniques like Wavelet Based Multi-Fractal Detrended Fluctuation Analysis (WB-MFDFA) can be extracted and probed for the identification of self-similarity at varying length scales due to the inherent use of fractal like kernels (for example Daubechies' family \cite{daubechies1992}).
\par
Wavelet transform due to it's multi-resolution analysis capability using the Daubechies' basis which extract the polynomial trends (for example, Db-4 and Db-6 extract the linear and quadratic trends respectively) has been shown to characterize the scaling behavior and self-similarity of empirical data sets quite faithfully \cite{mani2005,mani2009}. Indeed, it has been initially explored to analyze tissue fluorescence spectra in an attempt to distinguish between normal and dysplastic tissue \cite{gupta2005,agarwal2003,gharekhan2008,gharekhan2010,gharekhan2011}. In this work, we employ this multi-resolution property of wavelets to ascertain the changes in the self-similarity of dysplastic human cervical tissues as opposed to healthy human cervical tissues by analyzing the esoteric nature of the fluctuations in tissue light scattering spectra.
\par
This article is organized as follows: Fourier and power spectrum analysis is reviewed in \ref{sec:FTPLA}, discrete wavelet transform in \ref{sec:DWT}, wavelet based power law analysis in \ref{sec:WFPLA}, wavelet based multi-fractal de-trended fluctuation analysis in \ref{sec:WBMFDFA} and correlation based analysis in \ref{sec:CBA}. Sec. \ref{sec:EMM} describes the experimental methods for light scattering measurements from tissues. Sec. \ref{sec:obsres} deals with our findings from the analysis and contains a discussion of the same in the context of the differences between the normal and dysplastic samples. In Sec. \ref{sec:conclusions} we conclude with the prospect of pre-cancer detection using light scattering techniques combined with novel fluctuation analysis methods.
\section{Theory}
\label{sec:theory}
\subsection{Fourier analysis and power law spectrum}
\label{sec:FTPLA}
Fourier Analysis has traditionally been a preferred tool for analysis of experimental data sets. Here, we just briefly review the Discrete Fourier Transform (DFT) and its power spectrum. For a data set $x(n),n=\{1,2,\ldots\}$, the DFT is a linear transformation over an orthogonal basis given by:
\begin{equation}
x(k)=\sum_{n=0}^{N-1} x(n) \exp \left(-\frac{2 \pi \imath}{N} k n\right), \quad k \in [0,N-1]
\end{equation}
and it's power spectrum
\begin{equation}
P(k)=\Bigg\vert \sum_{n=0}^{N-1} x(n) \exp\left(-\frac{2 \pi \imath}{N} k n\right) \Bigg\vert^2
\end{equation}
is known to follow a power law behavior for self-similar processes
\begin{equation}
P(k)\approx k^{\alpha}.
\end{equation}
The power law coefficient $\alpha$ is related to the Hurst scaling exponent $H \in (0,1)$, and the fractal dimension $D_f$ which identify the nature of fractal behavior by $\alpha=2H+1=3-D_f$ \cite{hurst1951}. Though this analysis assumes a global scaling behavior, due to the presence of inhomogeneities in the intra-cellular structure, the scaling behavior turns out to be localized in the spatial frequency domain. In multi-fractal analysis, a local Hurst exponent is calculated which quantifies the local singular behavior and provides useful and minute information which is hidden between different scales. We therefore applied a more general wavelet based approach for the multi-fractal analysis.  
 
\subsection{A brief review of wavelet transforms}
\label{sec:DWT}
Wavelet Transforms \cite{daubechies1992,farge1992,torrence1998,mallat1989}, in both the forms of Discrete and Continuous transforms in the recent years have emerged as an invaluable tool in the field of data analysis and interpretation. Here, we will briefly review the Discrete Wavelet Transform (DWT).
\par
In DWT, two functions, namely, $\phi(n)$ and $\psi(n)$, called the father and the mother wavelets respectively, form the kernels. They satisfy the admissibility conditions:  $\int \phi(n) dn < \infty$, $\int \psi(n) dn = 0$, $\int \phi^*(n) \psi(n) dn =0$, $\int\vert \phi(n)\vert^2 dn = \int\vert \psi(n)\vert^2 dn=1$. The scaled and translated versions of the mother wavelet $\psi(n)$ are called the daughter wavelets
\begin{equation}
 \psi_{s,m}(n)=2^{s/2}\psi(2^s n-m),~~m \in \mathbb{R}, ~~s\in \mathbb{Z}^+,
\end{equation}
which form a complete set and differ from the former in terms of their height and width. At the $s^{th}$ scale, the height and width of the daughter wavelet are $2^s$ and $2^{s/2}$ of that of the mother wavelet respectively. $s$ and $m$ are the scaling and translation parameters. 

The DWT for a function $f(t)$ is given by,
\begin{equation}
 f(t)=\sum_{m=-\infty}^{\infty} a_m \phi_m (t) +\sum_{m=-\infty}^{\infty}\sum_{s\geq0}^{l}d_{s,m}\psi_{s,m}(t),
\end{equation}
the coefficient $a_m$ (called the approximation coefficient) extracts the trend and $d_{s,m}$ (called the detail coefficient) extracts the details or fluctuations from the signal. Here, $l=\lfloor \log(N)/ \log(2)\rfloor$ is the upper bound for taking the maximum scale for analysis above which the edge effects corrupt the wavelet coefficients giving rise to spurious results. This mathematical artifact is explained by the cone of influence \cite{torrence1998}.

\subsection{Wavelet Fluctuation Power Law Analysis}
\label{sec:WFPLA}
\begin{figure}[h]
\begin{centering}
\includegraphics[bb=180 435 450 726,clip=true,scale=0.6]{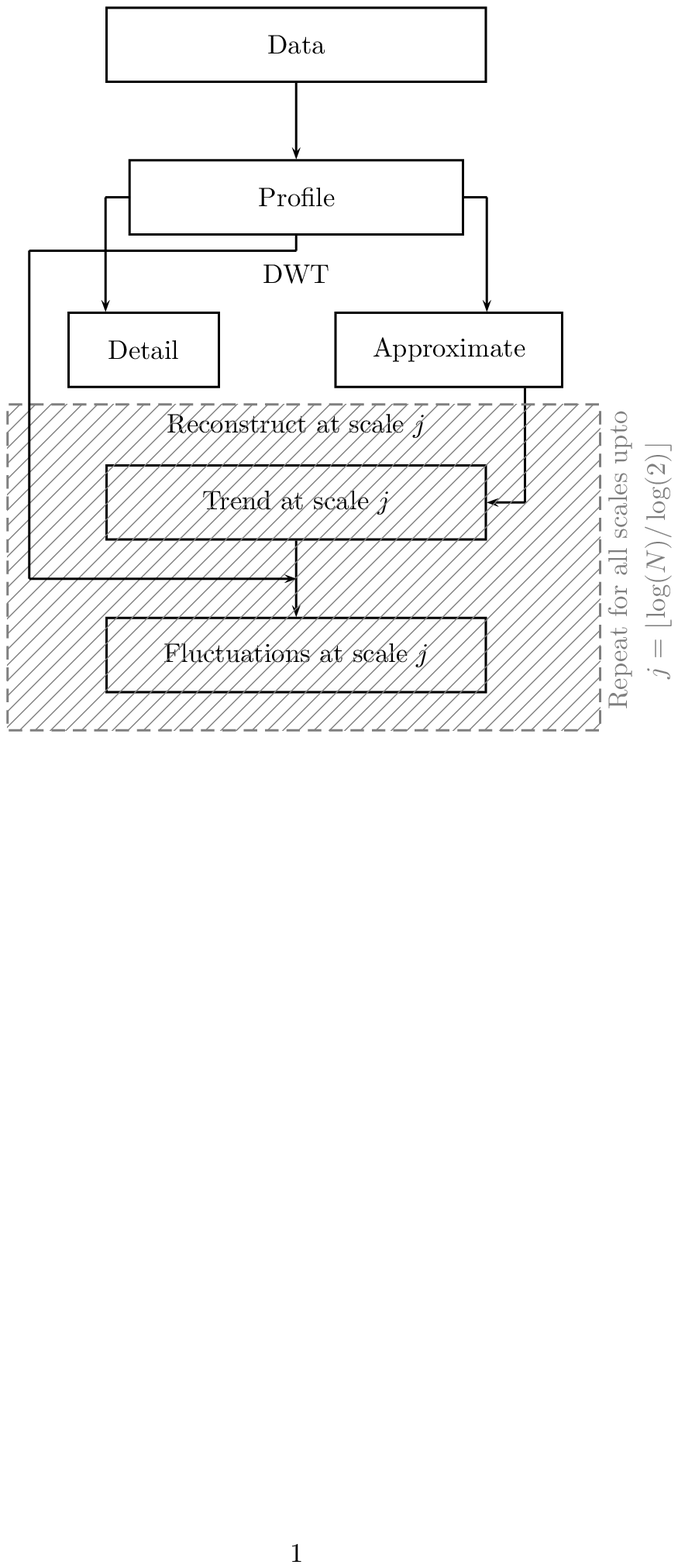}
\caption{\label{fig:algo} Schematic representation of the fluctuation extraction algorithm (adapted from \cite{sayantan2011}).}
\end{centering}
\end{figure}

The Daubechies' family of wavelets are made to satisfy vanishing moments conditions, and hence are able to isolate polynomial trends from a time series (see \cite{daubechies1992}). The different wavelets in this family isolate trends of different polynomial orders, for example, $Db-4$ isolates a monomial trend while $Db-6$ isolates a quadratic trend. We first obtain the profile from the fluctuations by taking their cumulative sum: $Y(n)=\sum_{m=1}^n X(m), n \in \{1,2,\ldots,N-1\}$, where, $Y$ and $X$ are profile and fluctuation signal respectively and $N$ is the data length. In this method, we first obtain the fluctuations at every scale by a wavelet reconstruction taking the approximation coefficients (using Db-4 wavelet) and subsequent subtraction from the signal. A flowchart for this fluctuation extraction is shown in Fig. \ref{fig:algo}.
 We then fit the Fourier power spectrum of these fluctuations to obtain the power law exponent as a function of scale, i.e. $\alpha \equiv \alpha(s)$ for the identification of the short and long term correlations. This technique sheds light on the scaling behavior of the fluctuations in different spatial frequency regimes. 
\subsection{Wavelet Based Multi Fractal De-trended Fluctuation Analysis}
\label{sec:WBMFDFA}
The Multi Fractal De-trended Fluctuation Analysis (WB-MFDFA) algorithm proposed by Manimaran \textit{et al.} \cite{mani2005,mani2009} has been employed gainfully to extract the multi-fractal nature of a variety of physical systems. In order to apply the WB-MFDFA algorithm, we then obtain the fluctuations using the extraction algorithm shown in Fig. \ref{fig:algo}. The asymmetric nature of the wavelet function and the edge-effects (due to the cone of influence) encountered during the convolution can affect the precision of the fluctuations. Hence, we repeat this procedure on a reversed profile and then take the average to get the fluctuations at every scale which are denoted by $F_{\tilde{s}}$. We used $\tilde{s}$ to represent the wavelet scale so as not to confuse with the segment length $s$ which is related to the wavelet scale $\tilde{s}$ by the number of filter coefficients for a given wavelet. These fluctuations are then segmented into $M_s$ non-overlapping sections such that $M_s=\lfloor N/s \rfloor$, where, $s$ and $N$ are the window size and the length of the fluctuations respectively. Subsequently, we obtain the $q^{th}$ order fluctuation function $F_q(s)$ by
\begin{subequations}
\begin{equation}
F_q(s)=\left[ \frac{1}{M_s} \sum_{m=1}^{2M_s}\left\{F^2(m,s)\right\}^{\frac{q}{2}}\right]^{\frac{1}{q}}, \qquad q\neq 0
\end{equation}
\begin{equation}
\mbox{and,}\quad F_{q=0}(s)=\exp \left[ \frac{1}{M_s} \sum_{m=1}^{2M_s}\log \left\{F^2(m,s)\right\}^{\frac{q}{2}}\right]^{\frac{1}{q}}, \qquad q=0.
\end{equation}
\end{subequations}
where the order of moments $q$ can and both positive and negative integers. The fluctuation function for self similar processes follows a scaling law, the scaling function given by $F_q(s)\approx s^{h(q)}$. We should note here that the smaller fluctuations in the light scattering spectrum will be influenced by the negative $q$ values, whereas the positive values of $q$ will impact the larger fluctuations. The scaling function $h(q)$ calculated at $q=2$ corresponds to the Hurst exponent \cite{hurst1951}. $H=0.5$ represents uncorrelated (white noise, $f^{0}$) or brown noise $f^{-2}$, while $H>0.5$ represents long range correlations or persistent behavior. $H<0.5$ reveals short range correlations or anti-persistent behavior. Mono-fractals are scale independent and hence their $h(q)$ is independent of $q$. They can be characterized by a single parameter like the fractal dimension. However, for multi-fractals, the $h(q)$ is not independent of $q$ and they require a more complex function like the singularity spectrum for its characterization \cite{mandelbrot1982}.
\par
The classical multi-fractal scaling exponent $\tau(q)$ defined by standard partition function based formalism \cite{eke2002,stanley1988} is related to the Hurst exponent h(q) by $\tau(q)=qh(q)-1$. The multi-fractality can also be characterized by using the singularity spectrum $f(\beta)$, which is related to $\tau(q)$ through a Legendre transform: $\beta=d/dq [\tau(q)]$ and $f(\beta)=q \beta-\tau(q)$; where $\beta$ is the singularity strength or H\"older Exponent \cite{kantelhardt2002}. $f(\beta)$ denotes the dimension of the subset of the series to be characterized. $\beta$, $f(\beta)$ and $h(q)$ are related by 
\begin{equation}
f(\beta)=q[\beta-h(q)]+1.
\label{eq:fbeta}
\end{equation} $\beta$ will have a constant value for mono-fractal series, whereas for multi-fractal series $\beta$ values will have a distribution. Just like in the case of $h(q)$, a constant $\tau(q)$ will signify a mono-fractal, while a multi-fractal will depend on the order of the moments. The singularity spectrum is broader when correlations are present in the series and if the correlations are absent or lost, then the singularity spectrum becomes narrower.
\subsection{Correlation based analysis}
\label{sec:CBA}
The correlation matrix analysis method has been extensively used to study various physical and biological systems (for example see \cite{seba2003}). For two variables $x_i \in X$ and $y_j \in Y$, the correlation matrix is given by:
\begin{equation}
C_{ij}=\frac{\mean{x_iy_j}-\mean{x_i}\mean{y_j}}{\sqrt{(\mean{x_i^2}-\mean{x_i}^2)(\mean{y_j^2}-\mean{y_j}^2)}}
\end{equation}
where, $\mean{\ldots}$ is the mean. It is well known that multi-fractality is related to the domain structure formation in the correlation matrices and we present the correlation matrix analysis in the wavelength domain to support the wavelet based multi-fractal de-trended fluctuation analysis.
\section{Experimental materials and methods}
\label{sec:EMM}
\begin{figure}[h]
\begin{centering}
\includegraphics[bb=145 320 500 620,clip=true,scale=0.65]{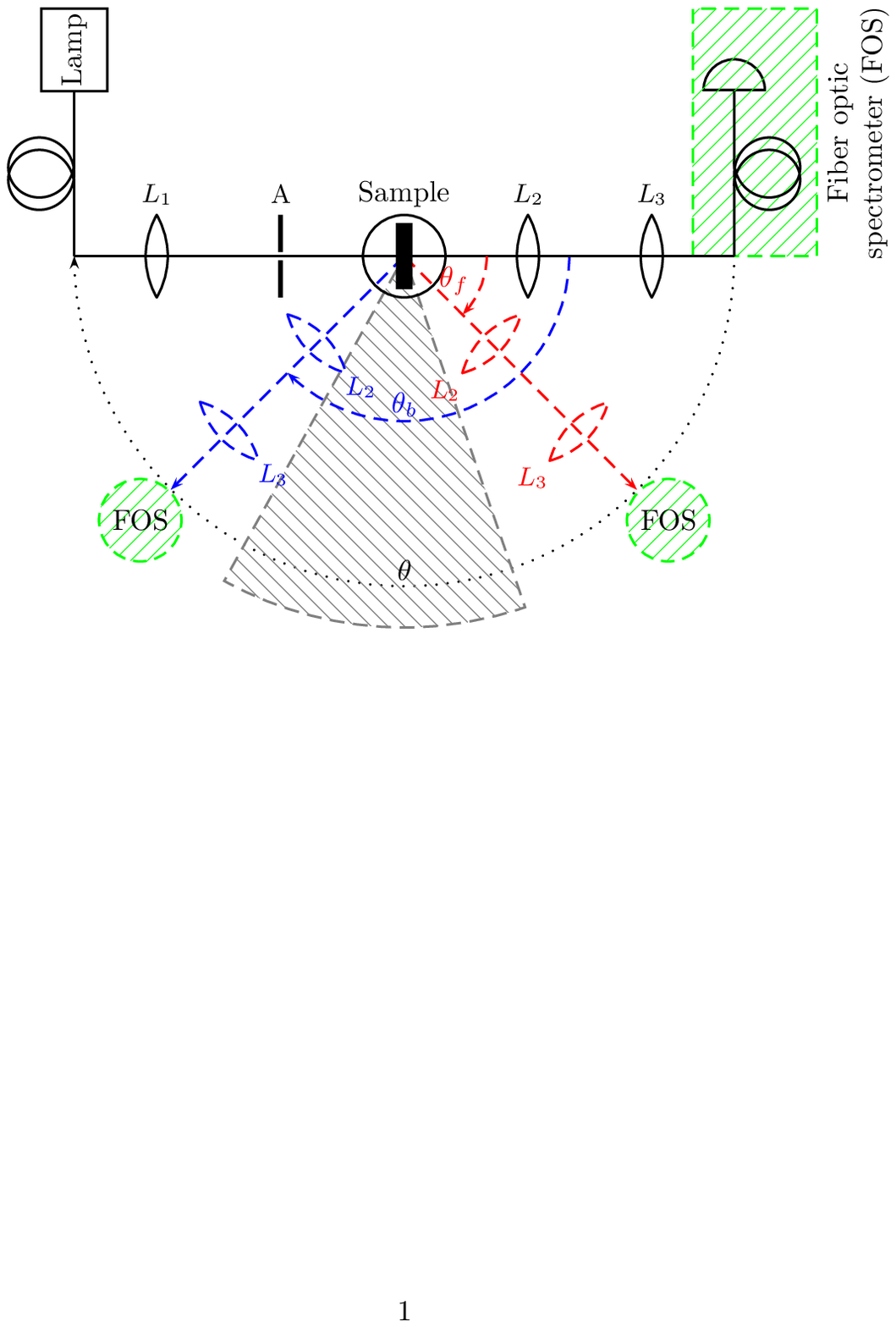}
\caption{\label{fig:schematic} (Color Online) Schematic representation of the experimental setup for the light scattering measurement. The sample mounted on a goniometer is illuminated by a white light collimated from a Xe-source and the scattered light is then collimated using two lenses ($L_2$ and $L_3$) and recorded using a Fiber-Optic Spectrometer. The $\theta_f$ represents the forward scattering angle, while $\theta_b$ is the backward scattering angle. The forward and backward scattering process are represented by red and blue lines respectively while the Fiber-Optic Spectrometer setup is shown in green. $0 \leq\theta\leq 180^\circ$ is the scattering angle at which the spectra were recorded. We must note in the gray shaded region extending from $80^\circ$ to $110^\circ$, the scattering spectra was not recorded due to geometry induced artifacts.}
\end{centering}
\end{figure}

Pathologically graded (CIN-I or dysplastic) unstained sections of human cervical tissue and their normal counterparts were frozen and vertically sectioned (thickness $\sim 5 \mu m$, lateral dimension $\sim 4mm \times 6mm$) on glass slides were used for light scattering measurement and analysis. A schematic representation of the experimental setup is shown in Fig. \ref{fig:schematic}.
The angle resolved elastic scattering spectra were recorded using a goniometric arrangement for the spectral range $\lambda:400nm-800nm$ and angular range $\theta: 10^\circ-150^\circ$ at $10^\circ$ intervals. White light output from a Xe-lamp (Newport USA, 50-5000 W) was collimated using a combination of lenses and were made incident on the sample kept at the center of the goniometer. The spot-size incident on the sample was controlled by a variable aperture ($\sim 1mm$). The scattered light from the sample was collimated by a pair of lenses and was then focused into a collecting fiber probe, the distal end of which was coupled to a spectrometer (Ocean Optics HR2000). The recorded spectra at each scattering angle was normalized by the lamp spectra.
\section{Results and Discussions}
\label{sec:obsres}
\begin{figure}[h]
\begin{centering}
\subfigure[Data at $10^{\circ}$.]{
\includegraphics[scale=0.3]{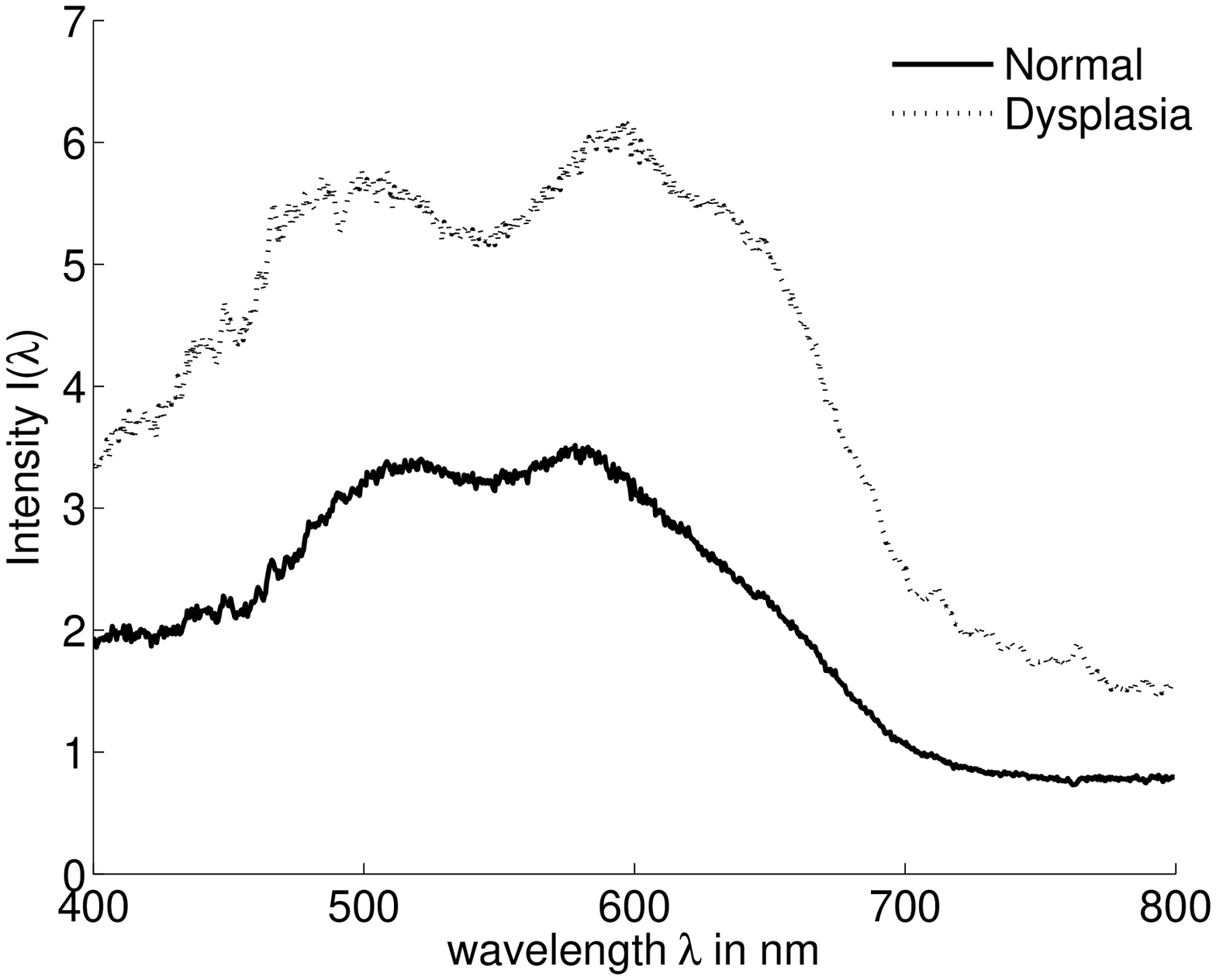}
\label{fig:data10}
}
\subfigure[Data at $150^{\circ}$.]{
\includegraphics[scale=0.3]{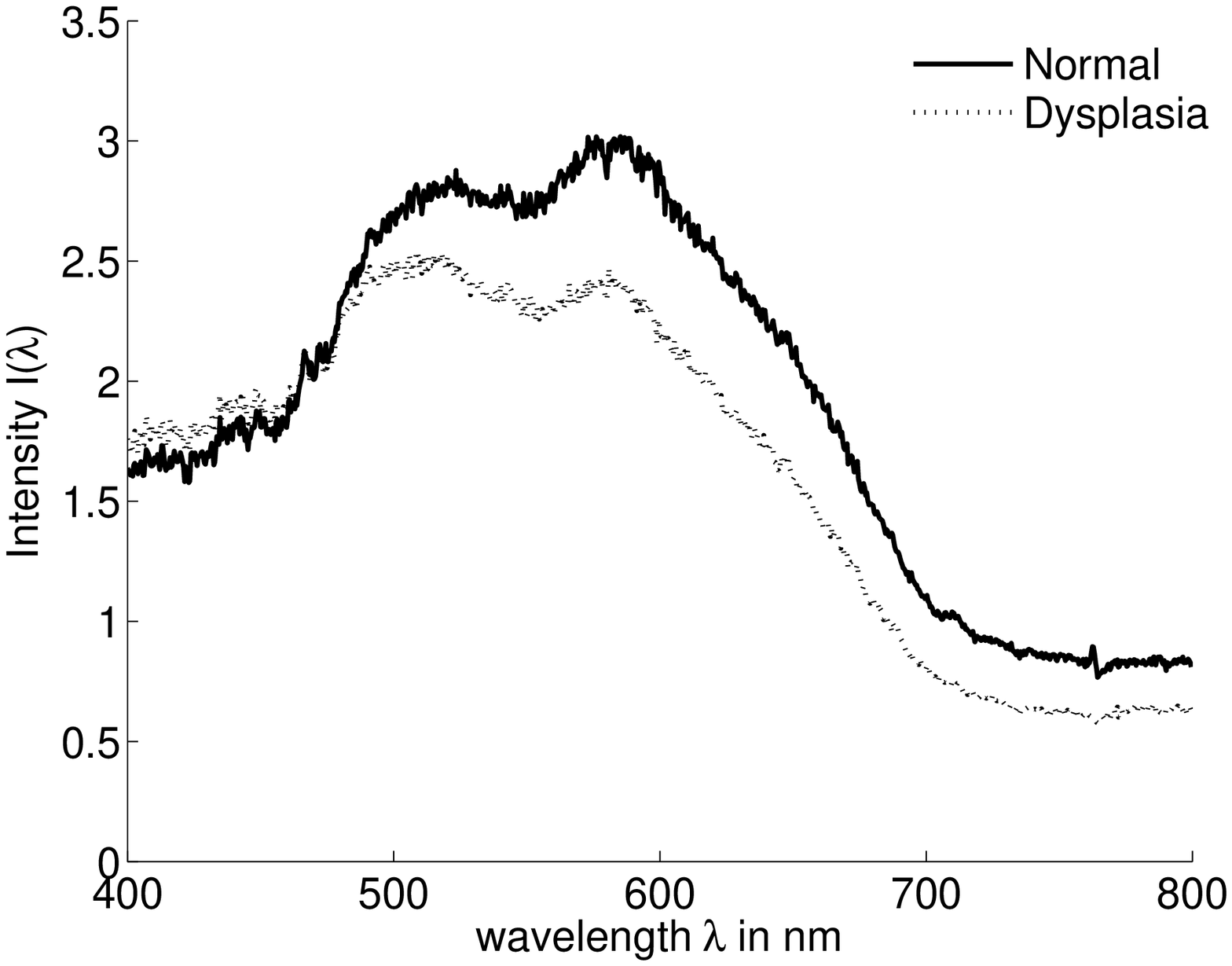}
\label{fig:data150}
}
\caption{\label{fig:data10150}Recorded elastic light scattering spectra at \subref{fig:data10} $10^{\circ}$ and \subref{fig:data150} $150^{\circ}$ from both normal and dysplastic tissues.}
\end{centering}
\end{figure}

Typical elastic light scattering spectra recorded from normal and dysplastic tissues are shown in Fig.\ref{fig:data10150}. Representative spectra for forward and back-scattering  are shown for $\theta=10^\circ$ and $\theta=150^\circ$ in Figs. \ref{fig:data10} and \subref{fig:data150} respectively. The broader structures are possibly due to larger sized scatterers, whereas the smaller features hidden in the spectra are the fluctuations due to the smaller sized scatterers \cite{gurjar2001}. We can see that the spectrum flattens as we move towards the backward scattering angles as compared to the forward scattering angles, which signifies the involvement of larger sized scatterers, such as the nucleus, in forward scattering and smaller sized intracellular organelles in the backward scattering. Such angular dependence of elastic scattering spectra from tissues have been extensively studied both experimentally and theoretically. These studies have indeed shown that the contribution of larger scatterers (like cells, nuclei) are more dominant in the spectra recorded at forward angles, whereas the spectra recorded at back-scattering angles are typically more influenced by the smaller sized scatterers \cite{drezek2003}.
\par
Although it is difficult to make a one-to-one correspondence between the intensity fluctuations and local refractive index fluctuations in the morphological structures, some information about the morphology can still be inferred from the light scattering spectral fluctuations. Indeed, it has been shown that with Born approximation, the light scattering spectrum can be represented as a Fourier transform power spectrum of the local refractive index fluctuations (see Sec. \ref{sec:FTPLA} for details). Thus, the observed spectral fluctuations contain information about the corresponding refractive index fluctuations in the Fourier domain as discussed below.
\par
\begin{figure}[ht]
\begin{centering}
\includegraphics[scale=0.4]{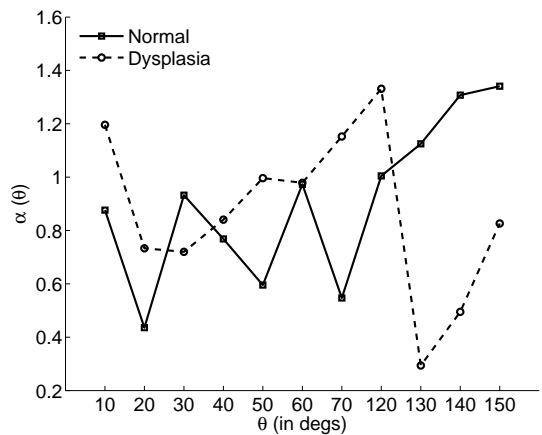}
\caption{\label{fig:AlphaFTWT}Variation of the scaling exponent $\alpha$ as a function of the scattering angle $\theta$ as obtained from the Fourier analysis.}
\end{centering}
\end{figure}

The results of the Fourier analysis on the light scattering spectra for different scattering angles is shown in Fig. \ref{fig:AlphaFTWT}. In the forward scattering region ($10^\circ-50^\circ$), we observe that the dysplastic samples show $\sim 0.7\leq\alpha\leq 1.2$, in the same region, however, the normal tissues show $0.4 \leq \alpha \leq 0.8$ (where $\alpha$ is the power law coefficient). The higher values of the power-law coefficient $\alpha$ corresponds to higher values of the Hurst scaling exponent $H$, indicative of ``coarseness'' of the fluctuations tending towards sub-fractal behavior; whereas at higher angles ($110^\circ-150^\circ$), $\alpha$ values are observed to be smaller in dysplastic samples than that for normal samples in the same region indicating more ``roughness'' in the fluctuations, signifying a trend towards extreme fractality.
\par
The higher $\alpha$ values in the angular range $(10^\circ \leq \theta \leq 70^\circ)$ for the dysplastic tissues as shown in Fig. \ref{fig:AlphaFTWT}, due to the more dominant contribution of the large scale scattering structures. This possibly can be related to the proliferative nuclear morphology in dysplasia \cite{gurjar2001}. This follows because such morphological changes are expected to lead to coarser fluctuations in light scattering spectra as is evident from the higher value of $H$ in the dysplastic tissues in the forward scattering range. On the other hand, the lower values of $\alpha$ in the back-scattering angles $(110^\circ \leq \theta \leq 150^\circ)$ indicates the predominance of small scale heterogeneity in the dysplastic tissues. It is also pertinent to note that around $60^\circ$, there is no distinct difference, possibly due to the simultaneous contribution of different sized scatterers in this angular range resulting in ``lumped'' effects.

\par
\begin{figure}[htb]
\begin{centering}
\subfigure[Normal]{
\includegraphics[scale=0.30]{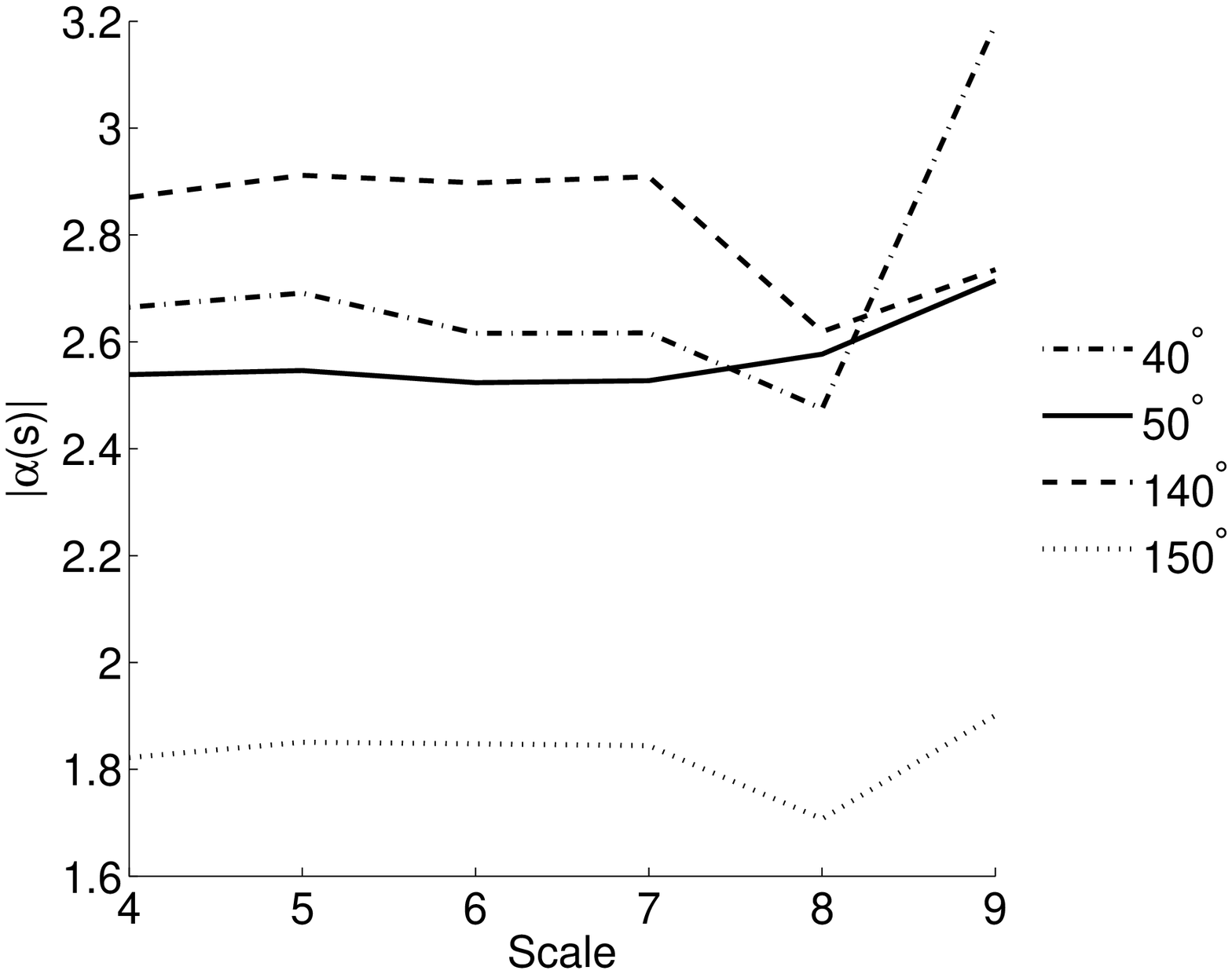}
\label{fig:NormalAlpha}
}
\subfigure[Dysplasia]{
\includegraphics[scale=0.30]{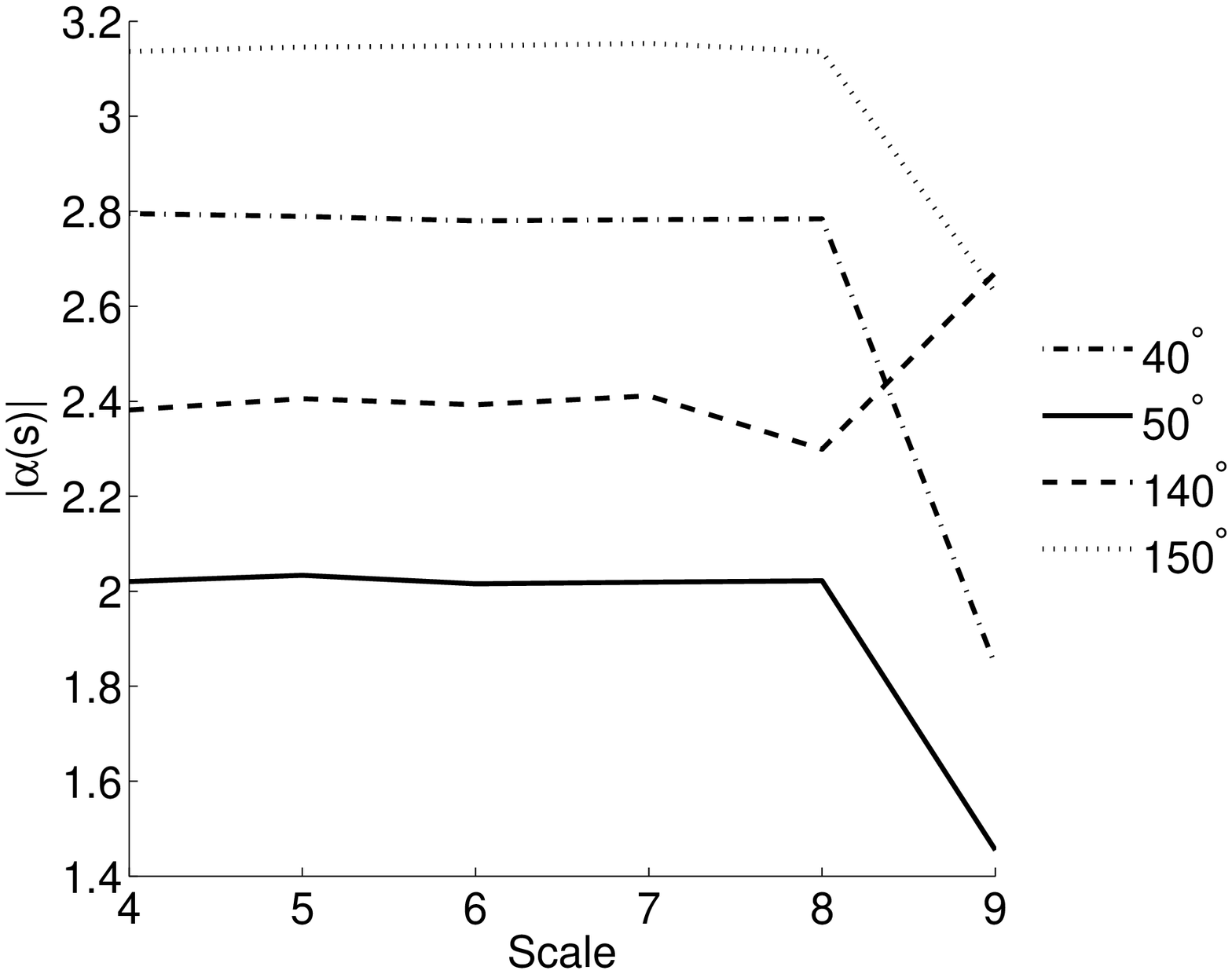}
\label{fig:CancerAlpha}
}
\caption{\label{fig:AlphaAll}Variation of the power law exponent $\alpha$ with the wavelet scale $s$ at representative forward and backward scattering angles for \subref{fig:NormalAlpha} normal and \subref{fig:CancerAlpha} dysplastic tissues.}
\end{centering}
\end{figure}

%\subsection{Results of WB-PLA}
In Fig. \ref{fig:AlphaAll}, we present the plots of power-law scaling exponent $\alpha$ as a function of the scale for normal (Fig. \ref{fig:NormalAlpha}) and dysplasia (Fig. \ref{fig:CancerAlpha}) for forward scattering angles $40^\circ$ and $50^\circ$ and for backward scattering angles $140^\circ$ and $150^\circ$. This analysis was performed following the discussion in Sec. \ref{sec:WFPLA}. It must be noted that the difference in the absolute value of $\alpha$ for Fourier and wavelet analysis arises from the fact that, in Fourier analysis, we analyze the signal itself while in wavelet analysis, we obtain the $\alpha$ values from the fluctuations which can be thought of as higher derivatives of the signal. We observe that the $\alpha$ values show a strong dependence on the scattering angle $\theta$. In both Fig. \ref{fig:NormalAlpha} and Fig. \ref{fig:CancerAlpha}, we observe the presence of a broad spectrum of processes such that $1.4\leq\alpha\leq3.2$. This arises possibly from the contribution of differently sized scatterers as has been mentioned earlier. We must however note that the behavior of the normal and dysplastic tissues in power law regime turn out to be very different. For example, in the forward scattering region, at $50^\circ$, we observe that the normal tissues show an average $\alpha=2.6$, while the dysplastic tissues show an average $\alpha=2.0$ over all scales. Similarly, in the backward scattering region, we obtain $\alpha=1.8$ and $\alpha=3.1$ for normal and dysplastic tissues respectively at $150^\circ$. In the other regions also, we find stark differences between normal and dysplastic samples in the $\alpha$ values averaged over all scales.
\par
\begin{figure}[h]
\begin{centering}
\includegraphics[scale=0.35]{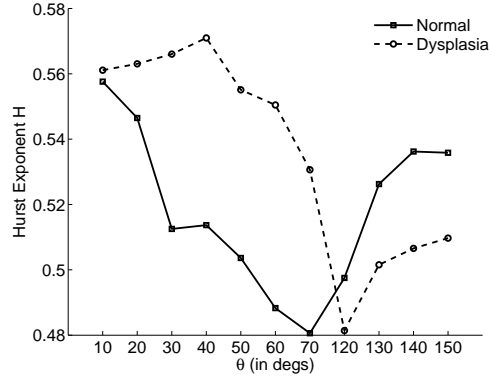}
\caption{\label{fig:hqtheta}The extracted Hurst parameter, $H=h(q=2)$ as a function of the scattering angle $\theta$ for normal and dysplastic tissues.}
\end{centering}
\end{figure}

As discussed in Sec. \ref{sec:WBMFDFA}, we have depicted the Hurst exponent as a function of the scattering angle $\theta$ in Fig. \ref{fig:hqtheta}; where we notice that though for both the normal and dysplastic tissues show a ``near-random'' behavior corresponding to $H=0.5$ ($H=0.5$ is indicative of Brownian behavior), we observe that the dysplastic samples show a marked departure from the normal samples in their trend.
 As observed for $\alpha(\theta)$ in Fig. \ref{fig:AlphaFTWT}, $H$ has a higher value in the forward scattering range $10^\circ\leq \theta \leq 70^\circ$ while the trend reverses in the backward scattering range $120^\circ \leq \theta \leq 150^\circ$.
\begin{figure}[h]
\begin{centering}
\subfigure[Normal]{
\includegraphics[scale=0.35]{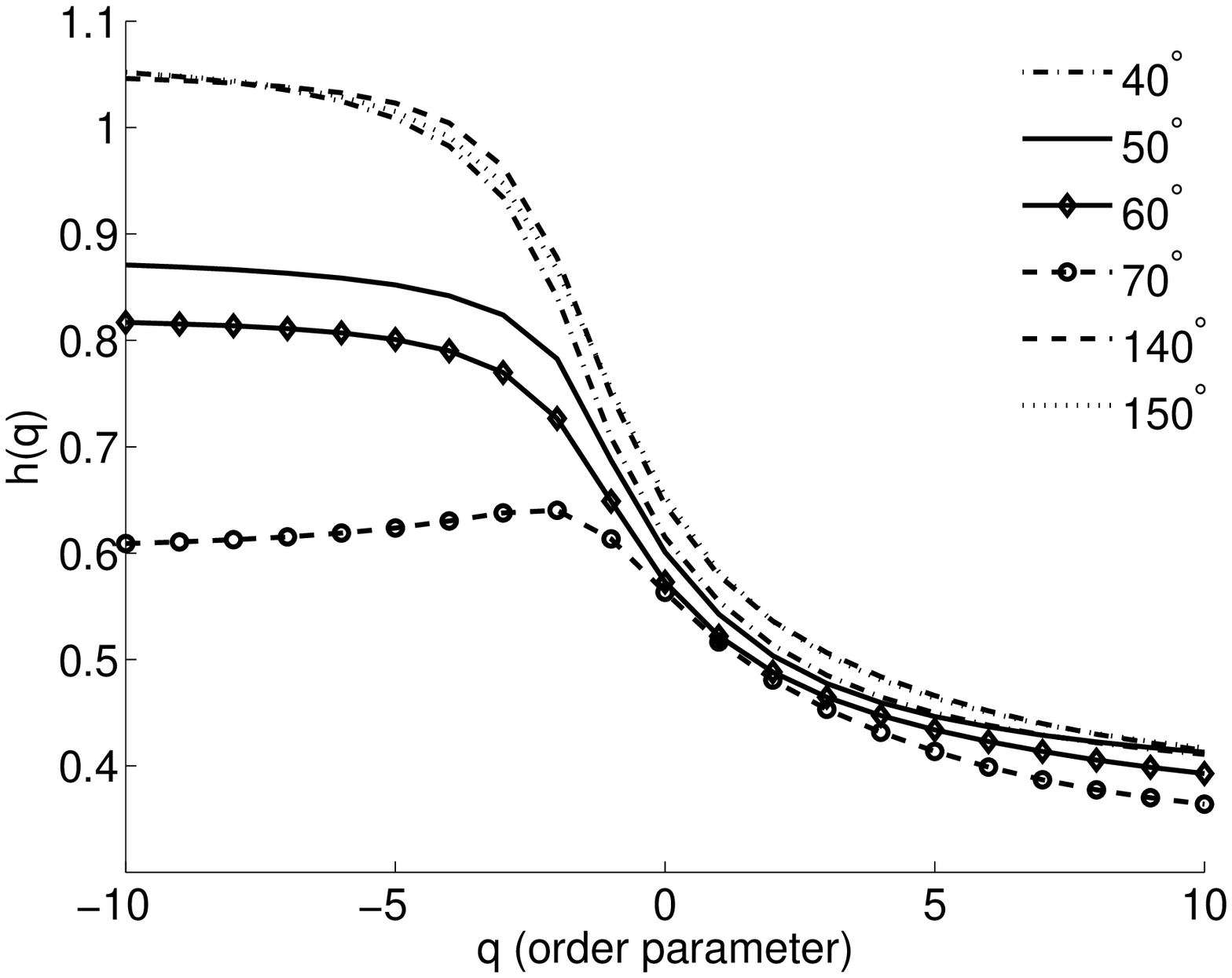}
\label{fig:normalhqq}
}
\subfigure[Dysplasia]{
\includegraphics[scale=0.35]{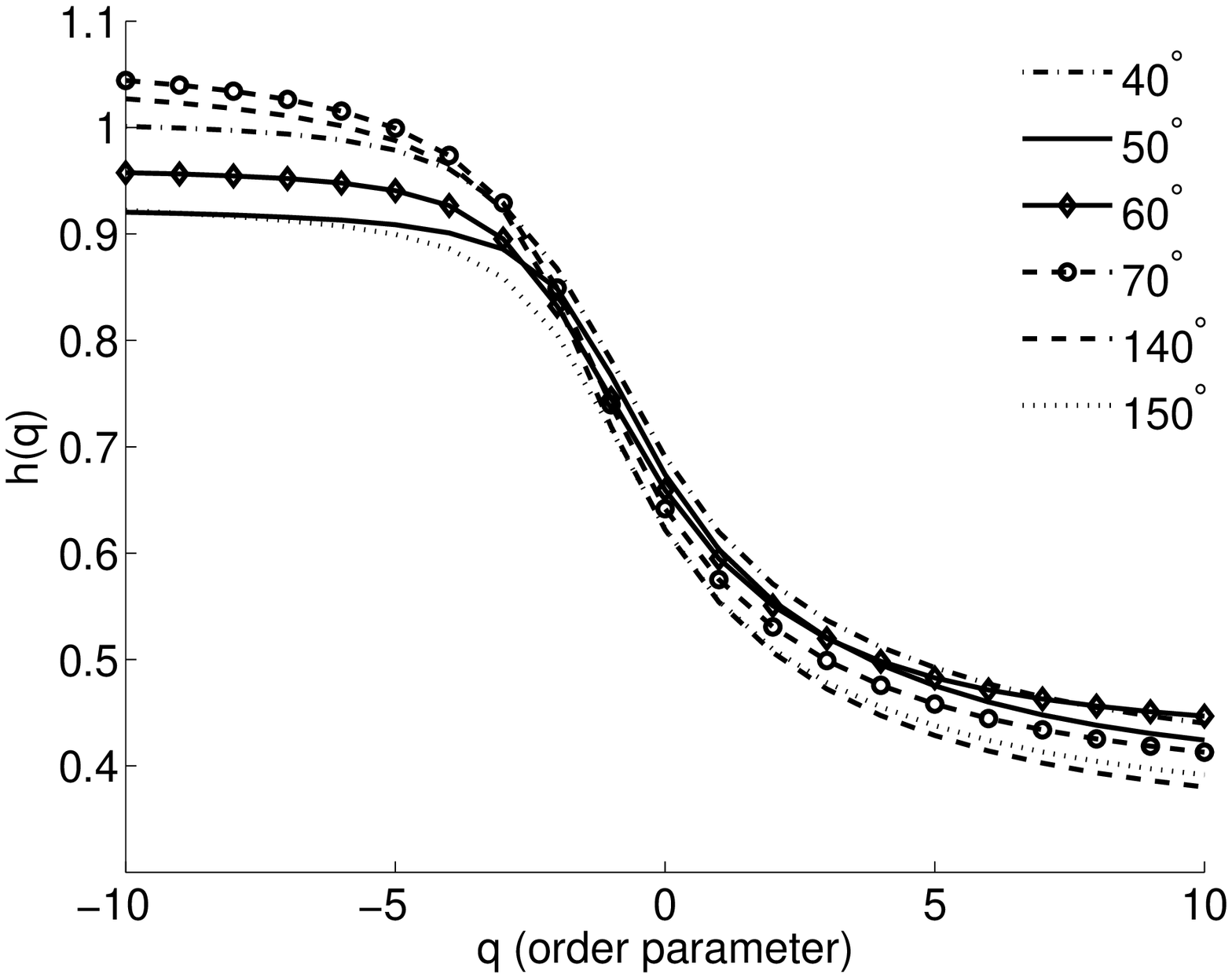}
\label{fig:cancerhqq}
}
\caption{\label{fig:hqq}The scaling function $h(q)$ at different forward and backward scattering angles for \subref{fig:normalhqq} normal and \subref{fig:cancerhqq} dysplasia. The weaker $q$ dependence of $h(q)$ for normal samples in the forward scattering angles $(50^\circ-70^\circ)$ is indicative of a possible mono-fractal trend, while the stronger dependence of the scaling function on the order of moments for dysplastic samples is indicative of a multi-fractal trend.}
\end{centering}
\end{figure}

\begin{figure}[ht]
\begin{centering}
\subfigure[Normal]{
\includegraphics[scale=0.35]{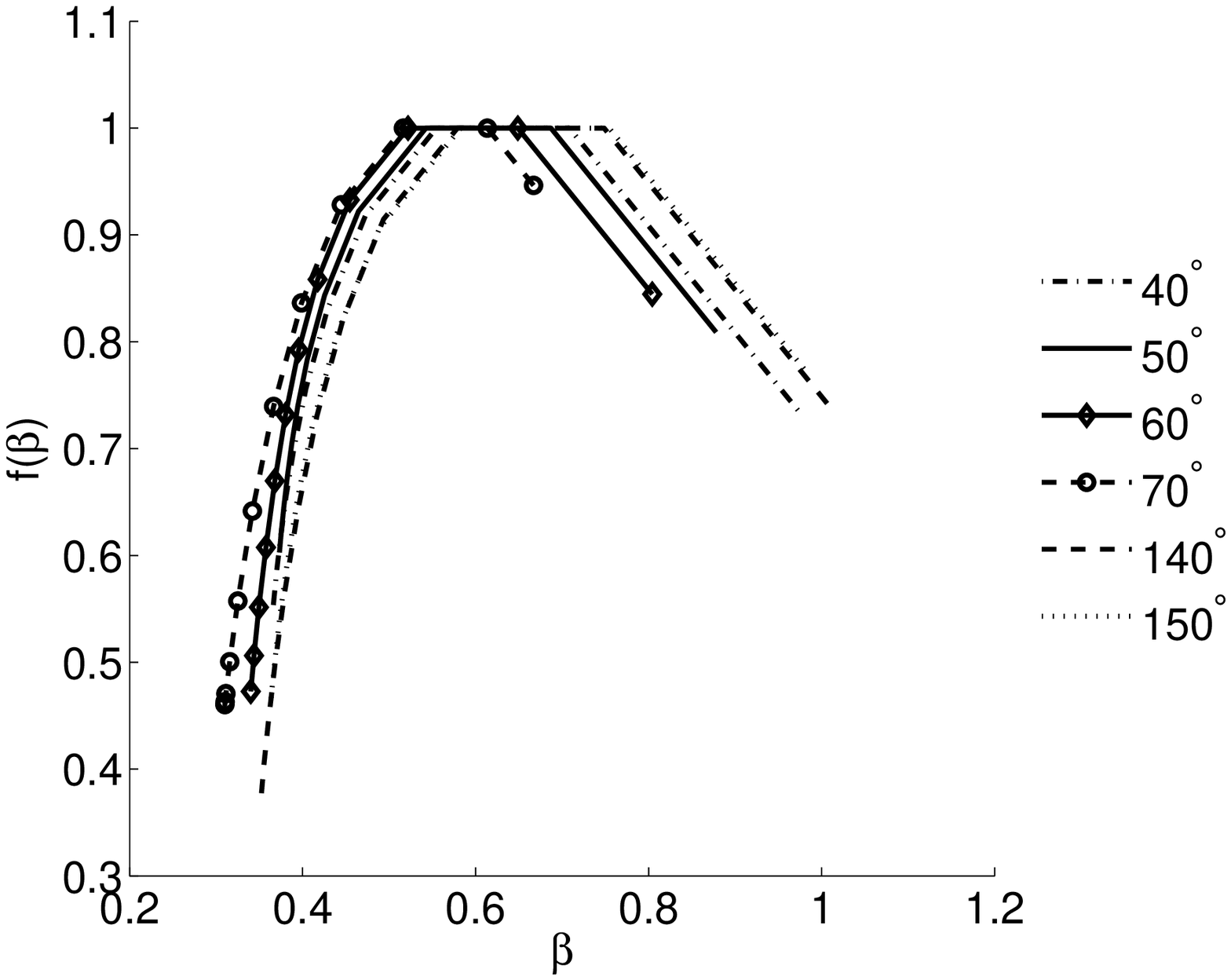}
\label{fig:NormalSingularity}
}
\subfigure[Dysplasia]{
\includegraphics[scale=0.35]{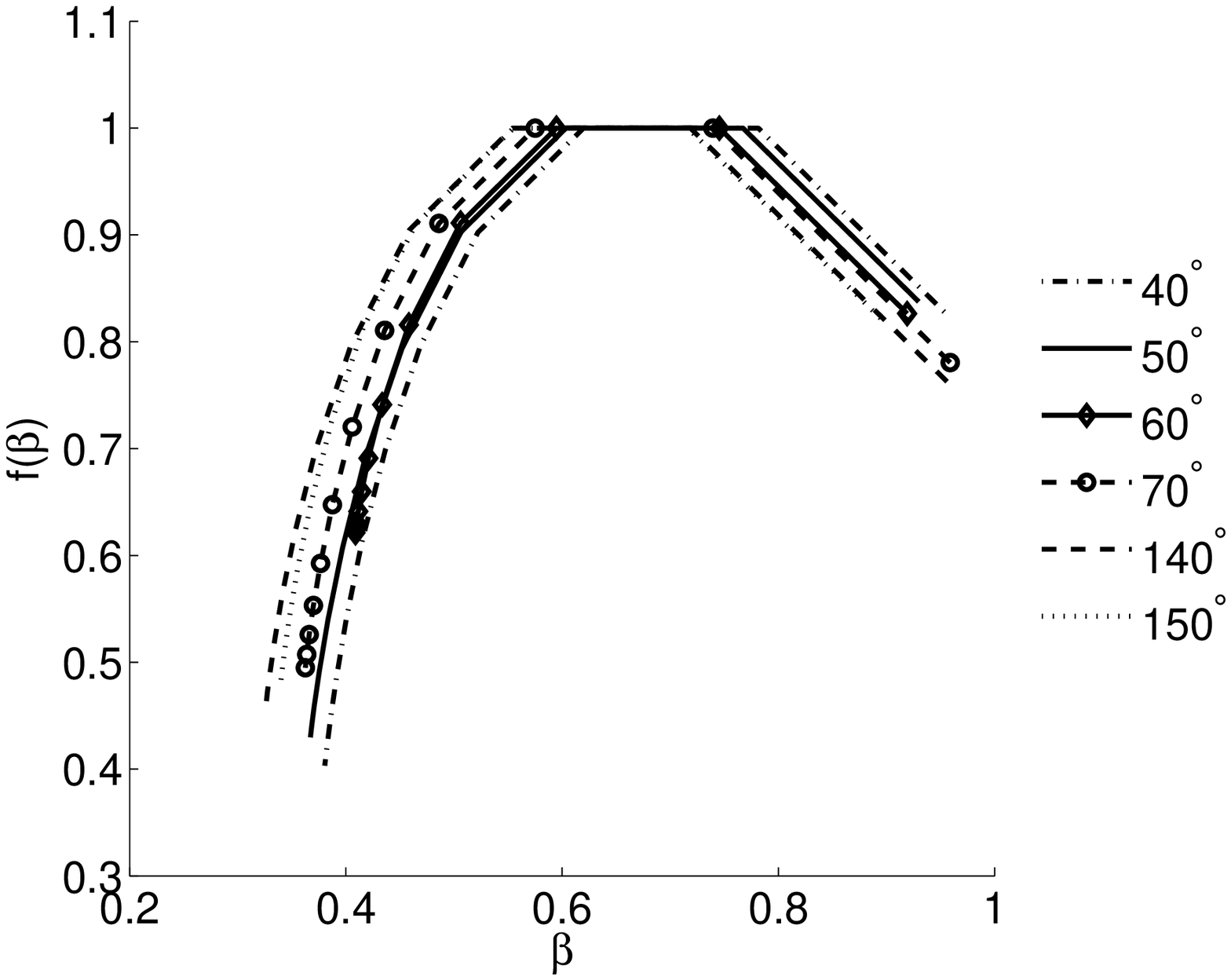}
\label{fig:CancerSingularity}
}
\caption{\label{fig:SingularityAll}The singularity spectrum $f(\beta)$ plotted against $\beta$ at all scattering angles $\theta$. \subref{fig:NormalSingularity} shows the singularity spectrum for normal samples while \subref{fig:CancerSingularity} shows the singularity spectrum for dysplastic tissues.}
\end{centering}
\end{figure}

The singularity spectrum $f(\beta)$ is a quantitative indicator of the exact nature of the self-similarity and its width represents the strength of the multi-fractality. In Fig. \ref{fig:SingularityAll},  we have shown the $f(\beta)$ as obtained from Eq. \ref{eq:fbeta} as discussed in Sec. \ref{sec:WBMFDFA}. We observe that the dysplastic samples have a higher multi-fractality than the normal samples, indicated by the width of singularity spectrum. It must be noted that for a mono-fractal, the singularity spectrum is similar to a Gaussian with a very small variance. This is consistent with our observations of Fig. \ref{fig:hqq}, where, the plots of $h(q)$ vs $q$ for normal and dysplastic tissues at a few representative forward and backward scattering angles are shown. For mono-fractals, the $h(q)$ is independent of $q$. In Fig. \ref{fig:hqq}, we observe that for the normal samples, at angles $50^\circ-70^\circ$, the dependence of $h(q)$ on $q$ is very small ($\Delta h(q)\sim 0.2$), which implies a trend towards mono-fractality. The normal samples thus show a wide range of variation from mono-fractals to multi-fractals. However, for dysplasia, the $h(q)$ dependence on $q$ is high ($\Delta h(q) \sim 0.7$ ) for all scattering angles, which indicates that the dysplastic samples show more multi-fractality than the normal samples.
\par
\begin{figure}[h]
\begin{centering}
\subfigure[Normal]{
\includegraphics[scale=0.30]{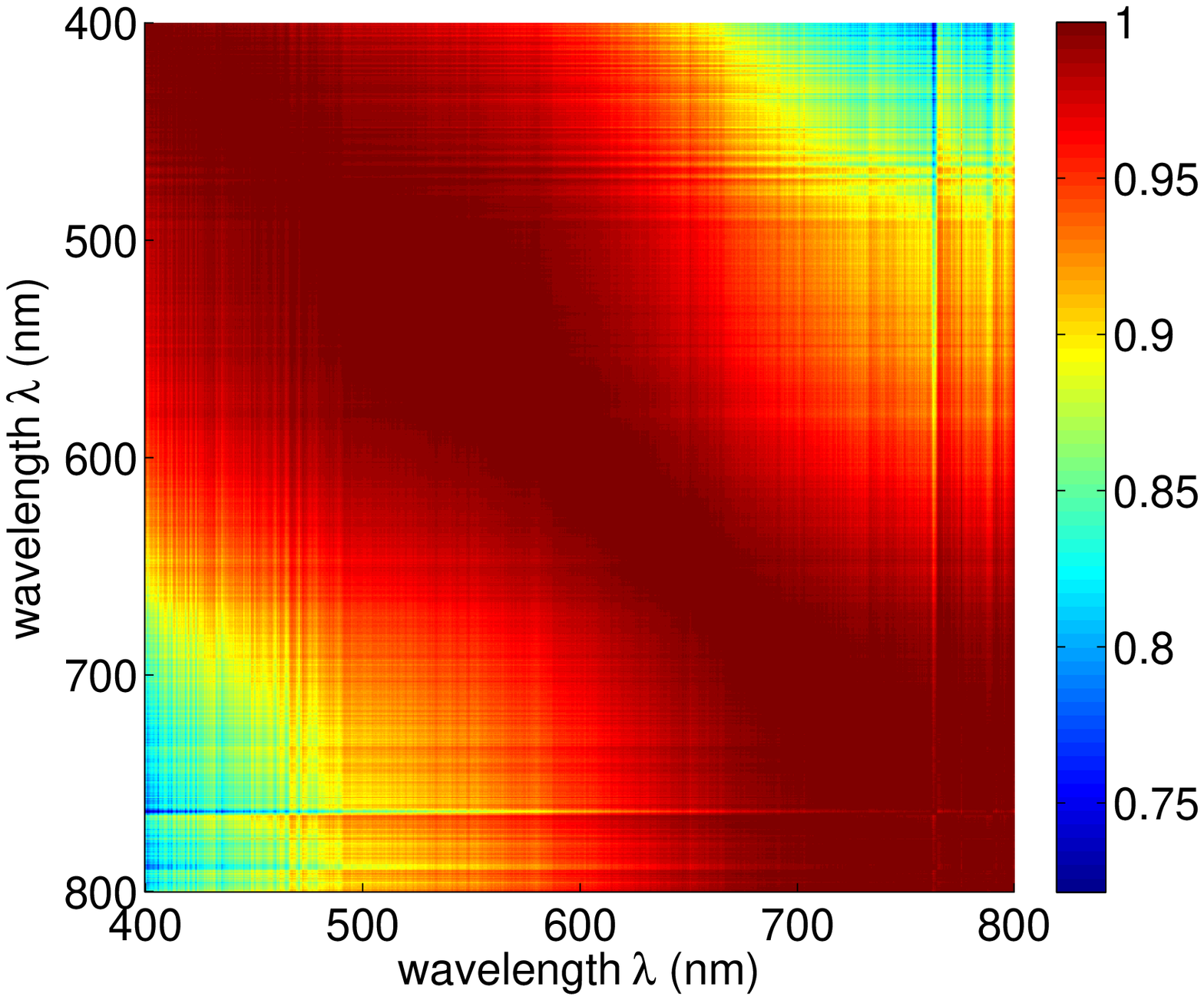}
\label{fig:mycorrnormal}
}
\subfigure[Dysplasia]{
\includegraphics[scale=0.30]{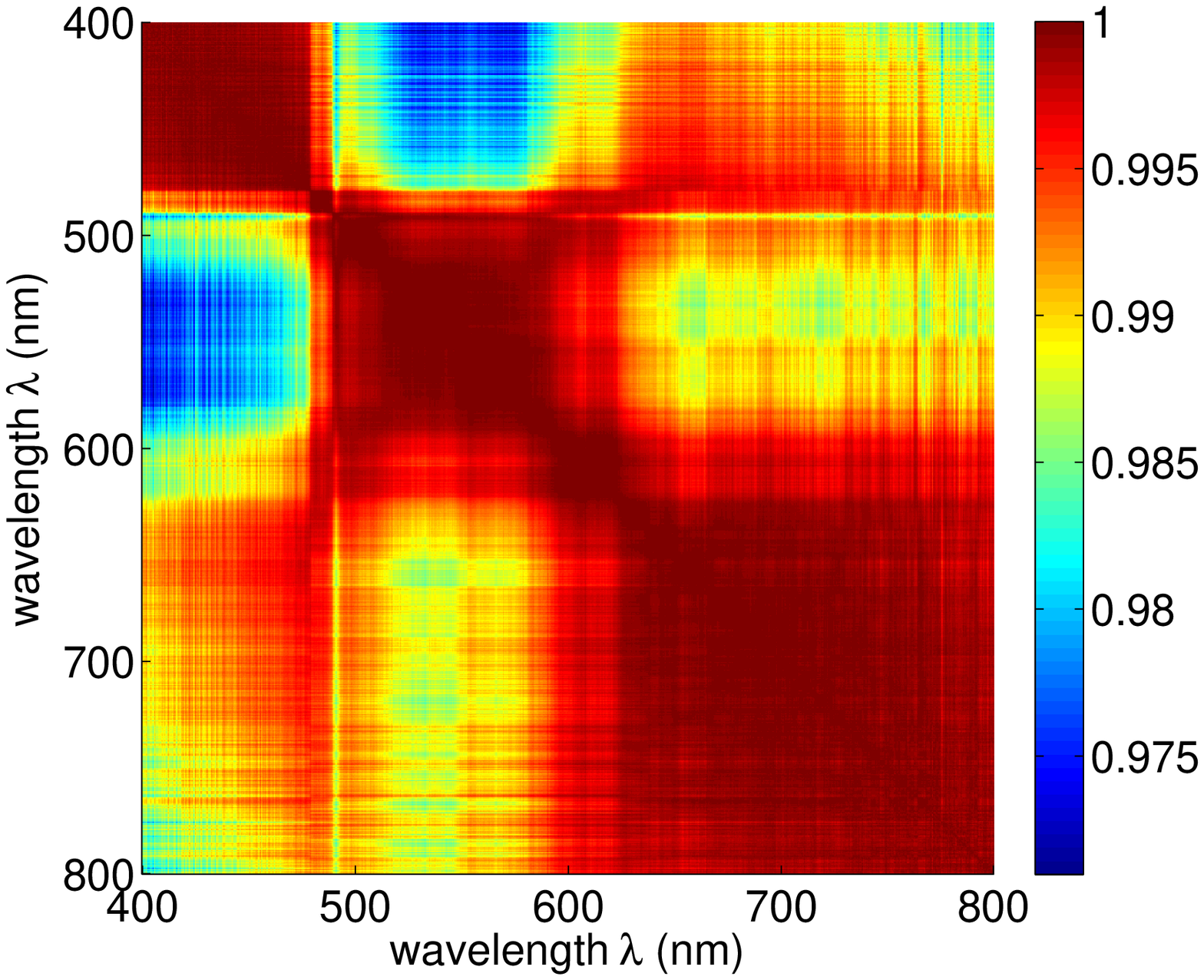}
\label{fig:mycorrcancer}
}
\caption{\label{fig:corr} (Color Online) The correlation matrices in the wavelength domain for \subref{fig:mycorrnormal} normal and \subref{fig:mycorrcancer} dysplastic samples.}
\end{centering}
\end{figure}

%\subsection{Correlations}
The spectral correlation matrices are shown in Fig. \ref{fig:corr} (following Sec. \ref{sec:CBA}). It is clear from Fig. \ref{fig:mycorrnormal} and \ref{fig:mycorrcancer} that though the normal tissues do not show any distinct correlation sectors in this domain, other than the expected correlation that would occur at neighboring wavelengths; dysplastic tissues show the presence of three dominant sectors. It must also be noted that the range of correlation increases from $0.70-1.00$ for normal to $0.97-1.00$ for dysplasia. This indicates a higher correlated behavior of the dysplastic tissues than the normal tissues in addition to domain formations in the spectral range. This possibly arises due to the fact that during dysplastic progression, the homogeneous cell morphology gives way to a more fragmented and heterogeneous structure.

\section{Conclusion}
\label{sec:conclusions}
In conclusion, we have applied a combined Fourier based and discrete wavelet based analysis on the fluctuations extracted from the elastic light scattering spectra of normal and dysplastic human cervical tissues. This approach clearly revealed otherwise hidden signatures of self-similarity in spectral fluctuation for both normal and dysplastic tissues, with significant differences in the nature of self-similarity. Fourier analysis of these fluctuations indicated the existence of multi-fractal nature and was further confirmed by WB-MFDFA. Dysplastic tissues showed marginally higher multi-fractality over the entire angular region compared to their normal counter-parts. The scaling exponent was observed to have angular dependence, possibly arising from the size distribution of the scatterers present in such complex systems. Note that this novel fluctuation analysis approach was initially explored on elastic scattering spectra recorded from a limited number of tissue samples, and the results were qualitatively similar. A more systematic study on statistically significant number of samples is currently underway, and the results will be reported in the near future. Never-the-less, early indications show promise of this approach for quantification of the morphological alterations associated with pre-cancers. The understanding gained from such analysis on the multi-fractal nature of fluctuations in tissues may ultimately lead to the development of non-invasive optical tools for pre-cancer detection.

\begin{thebibliography}{10}
\newcommand{\enquote}[1]{``#1''}

\bibitem{ramanujam2000}
N.~Ramanujam, \enquote{{Fluorescence spectroscopy of neoplastic and
  non-neoplastic tissues},} Neoplasia \textbf{2}, 89 (2000).

\bibitem{kortum1996}
R.~Richards-Kortum and E.~Sevick-Muraca, \enquote{{Quantitative optical
  spectroscopy for tissue diagnosis},} Annu. Rev. Phys. Chem. \textbf{47},
  555--606 (1996).

\bibitem{ghosh2005}
N.~Ghosh, S.~K. Majumder, H.~S. Patel, and P.~K. Gupta,
  \enquote{{Depth-resolved fluorescence measurement in a layered turbid
  mediumby polarized fluorescence spectroscopy},} Opt. Lett. \textbf{30},
  162--164 (2005).

\bibitem{ghosh2002}
N.~Ghosh, S.~K. Majumder, and P.~K. Gupta, \enquote{{Polarized fluorescence
  spectroscopy of human tissues},} Opt. Lett. \textbf{27}, 2007--2009 (2002).

\bibitem{haka2005}
A.~S. Haka, K.~E. Shafer-Peltier, M.~Fitzmaurice, J.~Crowe, R.~R. Dasari, and
  M.~S. Feld, \enquote{{Diagnosing breast cancer by using Raman spectroscopy},}
  Proc. Natl. Acad. Sci. (USA) \textbf{102}, 12371--12376 (2005).

\bibitem{boustany2010}
N.~N. Boustany, S.~A. Boppart, and V.~Backman, \enquote{{Microscopic Imaging
  and Spectroscopy with Scattered Light},} Annu. Rev. Biomed. Eng. \textbf{12},
  285--314 (2010).

\bibitem{fujimoto2003}
J.~Fujimoto, \enquote{{Optical coherence tomography for ultrahigh resolution in
  vivo imaging},} Nature Biotechnology \textbf{21}, 1361--1367 (2003).

\bibitem{schmitt1999}
J.~Schmitt, \enquote{{Optical coherence tomography (OCT): a review},} IEEE J.
  Sel. Topics Quantum Electron. \textbf{5}, 1205--1215 (1999).

\bibitem{hebden1997}
J.~C. Hebden, S.~R. Arridge, and D.~T. Delpy, \enquote{{Optical imaging in
  medicine: I. Experimental techniques},} Phys. Med. Biol. \textbf{42}, 825
  (1997).

\bibitem{ghosh2011}
N.~Ghosh, A.~Banerjee, and J.~Soni, \enquote{{Turbid medium polarimetry in
  biomedical imaging and diagnosis},} Eur. Phys. J. Appl. Phys. \textbf{54},
  {30001} (2011).

\bibitem{jacques2002}
{Jacques, S. L. and Ramella-Roman, J. C. and Lee, K.}, \enquote{{Imaging skin
  pathology with polarized light},} J. Biomed. Opt. \textbf{7}, 329--340
  (2002).

\bibitem{choi2007}
W.~Choi, C.~Fang-Yen, K.~Badizadegan, S.~Oh, N.~Lue, R.~R. Dasari, and M.~S.
  Feld, \enquote{{Tomographic phase microscopy},} Nature Methods \textbf{4},
  717--719 (2007).

\bibitem{gurjar2001}
R.~S. Gurjar, V.~Backman, L.~T. Perelman, I.~Georgakoudi, K.~Badizadegan,
  I.~Itzkan, R.~R. Dasari, and M.~S. Feld, \enquote{{Imaging human epithelial
  properties with polarized light-scattering spectroscopy},} Nature Medicine
  \textbf{7}, 1245--1248 (2001).

\bibitem{kalashnikov2009}
M.~Kalashnikov, W.~Choi, C.-C. Yu, Y.~Sung, R.~R. Dasari, K.~Badizadegan, and
  M.~S. Feld, \enquote{{Assessing light scattering of intracellular organelles
  in single intact living cells},} Opt. Express \textbf{17}, 19674--19681
  (2009).

\bibitem{choi2008}
W.~Choi, C.-C. Yu, C.~Fang-Yen, K.~Badizadegan, R.~R. Dasari, and M.~S. Feld,
  \enquote{{Field-based angle-resolved light-scattering study of single live
  cells},} Opt. Lett. \textbf{33}, 1596--1598 (2008).

\bibitem{graf2005}
R.~Graf and A.~Wax, \enquote{{Nuclear morphology measurements using Fourier
  domain low coherence interferometry},} Opt. Express \textbf{13}, 4693--4698
  (2005).

\bibitem{wax2003}
A.~Wax, C.~Yang, and J.~A. Izatt, \enquote{{Fourier-domain low-coherence
  interferometry for light-scattering spectroscopy},} Opt. Lett. \textbf{28},
  1230--1232 (2003).

\bibitem{perelman1998}
L.~T. Perelman, V.~Backman, M.~Wallace, G.~Zonios, R.~Manoharan, A.~Nusrat,
  S.~Shields, M.~Seiler, C.~Lima, T.~Hamano, I.~Itzkan, J.~Van~Dam, J.~M.
  Crawford, and M.~S. Feld, \enquote{{Observation of Periodic Fine Structure in
  Reflectance from Biological Tissue: A New Technique for Measuring Nuclear
  Size Distribution},} Phys. Rev. Lett. \textbf{80}, 627--630 (1998).

\bibitem{ghosh2010}
N.~Ghosh, M.~Wood, and A.~Vitkin, \enquote{{Polarized Light Assessment of
  Complex Turbid Media Such as Biological Tissues Using Mueller Matrix
  Decomposition},} in \enquote{Handbook of Photonics for Biomedical Science,} ,
  V.~V.~. Tuchin, ed. (CRC Press, 2010), Medical Physics and Biomedical
  Engineering, pp. 253--282.

\bibitem{tuchin2006}
V.~V. Tuchin, L.~Wang, and D.~A. Zimnyakov, \emph{Optical Polarization in
  Biomedical Applications} ({Springer-Verlag}, 2006).

\bibitem{ghosh2006}
N.~Ghosh, P.~Buddhiwant, A.~Uppal, S.~K. Majumder, H.~S. Patel, and P.~K.
  Gupta, \enquote{{Simultaneous determination of size and refractive index of
  red blood cells by light scattering measurements},} Appl. Phys. Lett.
  \textbf{88}, 084101 (2006).

\bibitem{ghosh2001}
N.~Ghosh, S.~K. Mohanty, S.~K. Majumder, and P.~K. Gupta, \enquote{{Measurement
  of Optical Transport Properties of Normal and Malignant Human Breast
  Tissue},} Appl. Opt. \textbf{40}, 176--184 (2001).

\bibitem{kim2006}
{Young L. Kim and Vladimir M. Turzhitsky and Yang Liu and Hariharan Subramanian
  and Prabhakar Pradhan}, \enquote{{Low-coherence enhanced backscattering:
  review of principles and applications for colon cancer screening },} J.
  Biomed. Opt. \textbf{11}, 041125 (2006).

\bibitem{drezek2003}
R.~Drezek, M.~Guillaud, T.~Collier, I.~Boiko, A.~Malpica, C.~Macaulay,
  M.~Follen, and R.~Richards-Kortum, \enquote{{Light scattering from cervical
  cells throughout neoplastic progression: influence of nuclear morphology, DNA
  content, and chromatin texture},} J. Biomed. Opt. \textbf{8}, 7 (2003).

\bibitem{yu2008}
C.-C. Yu, C.~Lau, G.~O'Donoghue, J.~Mirkovic, S.~McGee, L.~Galindo,
  A.~Elackattu, E.~Stier, G.~Grillone, K.~Badizadegan, R.~R. Dasari, and M.~S.
  Feld, \enquote{{Quantitative spectroscopic imaging for non-invasive early
  cancer detection},} Opt. Express \textbf{16}, 16227--16239 (2008).

\bibitem{capoglu2009}
\.{I}lker R.~\c{C}apo\u{g}lu, J.~D. Rogers, A.~Taflove, and V.~Backman,
  \enquote{{Accuracy of the Born approximation in calculating the scattering
  coefficient of biological continuous random media},} Opt. Lett. \textbf{34},
  2679--2681 (2009).

\bibitem{hunter2006}
M.~Hunter, V.~Backman, G.~Popescu, M.~Kalashnikov, C.~W. Boone, A.~Wax,
  V.~Gopal, K.~Badizadegan, G.~D. Stoner, and M.~S. Feld, \enquote{{Tissue
  Self-Affinity and Polarized Light Scattering in the Born Approximation: A New
  Model for Precancer Detection},} Phys. Rev. Lett. \textbf{97}, 138102 (2006).

\bibitem{xu2005}
M.~Xu and R.~R. Alfano, \enquote{{Fractal mechanisms of light scattering in
  biological tissue and cells},} Opt. Lett. \textbf{30}, 3051--3053 (2005).

\bibitem{sheppard2007}
C.~J.~R. Sheppard, \enquote{{Fractal model of light scattering in biological
  tissue and cells},} Opt. Lett. \textbf{32}, 142--144 (2007).

\bibitem{wu2007}
T.~T. Wu, J.~Y. Qu, and M.~Xu, \enquote{{Unified Mie and fractal scattering by
  biological cells and subcellular structures},} Opt. Lett. \textbf{32},
  2324--2326 (2007).

\bibitem{gao2010}
W.~Gao, \enquote{{Square law between spatial frequency of spatial correlation
  function of scattering potential of tissue and spectrum of scattered light},}
  J. Biomed. Opt. \textbf{15}, 030502 (2010).

\bibitem{wax2003_1}
A.~Wax, C.~Yang, M.~G. Müller, R.~Nines, C.~W. Boone, V.~E. Steele, G.~D.
  Stoner, R.~R. Dasari, and M.~S. Feld, \enquote{{In Situ Detection of
  Neoplastic Transformation and Chemopreventive Effects in Rat Esophagus
  Epithelium Using Angle-resolved Low-coherence Interferometry},} Cancer Res.
  \textbf{63}, 3556--3559 (2003).

\bibitem{schmitt1996}
J.~M. Schmitt and G.~Kumar, \enquote{{Turbulent nature of refractive-index
  variations in biological tissue},} Opt. Lett. \textbf{21}, 1310--1312 (1996).

\bibitem{perelman2006}
L.~Perelman, \enquote{Optical diagnostic technology based on light scattering
  spectroscopy for early cancer detection,} Expert Rev. Med. Devic. \textbf{3},
  787--803 (2006).

\bibitem{hurst1951}
H.~Hurst, \enquote{Long-term storage capacity of reservoirs,} Trans. Am. Soc.
  Civ. Eng. \textbf{116}, 770--808 (1951).

\bibitem{mandelbrot1982}
B.~Mandelbrot, \emph{The fractal geometry of nature} (W.H. Freeman, 1982).

\bibitem{kantelhardt2002}
J.~W. Kantelhardt, S.~A. Zschiegner, E.~Koscielny-Bunde, S.~Havlin, A.~Bunde,
  and H.~E. Stanley, \enquote{Multifractal detrended fluctuation analysis of
  nonstationary time series,} Physica A \textbf{316}, 87 -- 114 (2002).

\bibitem{mani2005}
P.~Manimaran, P.~K. Panigrahi, and J.~C. Parikh, \enquote{{Wavelet analysis and
  scaling properties of time series},} Phys. Rev. E \textbf{72}, 046120 (2005).

\bibitem{mani2009}
P.~Manimaran, P.~Panigrahi, and J.~Parikh, \enquote{{Multiresolution analysis
  of fluctuations in non-stationary time series through discrete wavelets},}
  Physica A: Statistical Mechanics and its Applications \textbf{388},
  2306--2314 (2009).

\bibitem{gupta2005}
S.~Gupta, M.~Nair, A.~Pradhan, N.~Biswal, N.~Agarwal, A.~Agarwal, and
  P.~Panigrahi, \enquote{{Wavelet-based characterization of spectral
  fluctuations in normal, benign, and cancerous human breast tissues},} J.
  Biomed. Opt. \textbf{10}, 054012 (2005).

\bibitem{agarwal2003}
N.~Agarwal, S.~Gupta, A.~Pradhan, K.~Vishwanathan, and P.~Panigrahi,
  \enquote{{Wavelet transform of breast tissue fluorescence spectra: a
  technique for diagnosis of tumors},} IEEE J. Sel. Topics Quantum Electron.
  \textbf{9}, 154--161 (2003).

\bibitem{gharekhan2008}
A.~Gharekhan, S.~Arora, K.~Mayya, P.~Panigrahi, M.~Sureshkumar, and A.~Pradhan,
  \enquote{{Characterizing breast cancer tissues through the spectral
  correlation properties of polarized fluorescence},} J. Biomed. Opt.
  \textbf{13}, 054063 (2008).

\bibitem{gharekhan2010}
A.~Gharekhan, S.~Arora, P.~Panigrahi, and A.~Pradhan, \enquote{{Distinguishing
  Cancer and Normal Breast Tissue Autofluorescence Using Continuous Wavelet
  Transform},} IEEE J. Sel. Topics Quantum Electron. \textbf{16}, 893--899
  (2010).

\bibitem{gharekhan2011}
A.~H. Gharekhan, S.~Arora, A.~N. Oza, M.~B. Sureshkumar, A.~Pradhan, and P.~K.
  Panigrahi, \enquote{{},} J. Biomed. Opt.  (2011). In Press.

\bibitem{daubechies1992}
I.~Daubechies, \emph{{Ten Lectures on Wavelets}}, {CBMS-NSF Regional Conference
  Series in Applied Mathematics} ({SIAM: Society for Industrial and Applied
  Mathematics}, 1992), 1st ed.

\bibitem{farge1992}
M.~Farge, \enquote{{Wavelet Transforms and their Applications to Turbulence},}
  Annu. Rev. of Fluid Mech. \textbf{24}, 395--458 (1992).

\bibitem{torrence1998}
C.~Torrence and G.~Compo, \enquote{{A practical guide to wavelet analysis},}
  Bull. Amer. Meteor. Soc. \textbf{79}, 61--78 (1998).

\bibitem{mallat1989}
S.~Mallat \emph{et~al.}, \enquote{{A theory for multiresolution signal
  decomposition: The wavelet representation},} IEEE Trans. Pattern Anal. Mach.
  Intell. \textbf{11}, 674--693 (1989).

\bibitem{sayantan2011}
S.~{Ghosh}, P.~{Manimaran}, and P.~K. {Panigrahi}, \enquote{{Characterizing
  Multi-Scale Self-Similar Behavior and Non-Statistical Properties of Financial

  Time Series},} Physica A  (2011). In Press.

\bibitem{eke2002}
A.~Eke, P.~Herman, L.~Kocsis, and L.~R. Kozak, \enquote{Fractal
  characterization of complexity in temporal physiological signals,} Physiol.
  Meas. \textbf{23}, R1 (2002).

\bibitem{stanley1988}
H.~E. Stanley and P.~Meakin, \enquote{{Multifractal phenomena in physics and
  chemistry},} Nature \textbf{335}, 405--409 (1988).

\bibitem{seba2003}
P.~\ifmmode~\check{S}\else \v{S}\fi{}eba, \enquote{{Random Matrix Analysis of
  Human EEG Data},} Phys. Rev. Lett. \textbf{91}, 198104 (2003).

\end{thebibliography}
\end{document}